\documentclass[aps,preprint,showpacs,showkeys]{revtex4}
\usepackage[pdftex]{graphicx}
\usepackage{amssymb}
\usepackage{amsmath}
\usepackage{bm}

\newcommand{\re}{\mathop{\rm Re}} \newcommand{\im}{\mathop{\rm Im}}
\newcommand\mS{m_{\text{S}}}
\newcommand\me{m_{\text{e}}}

\newcommand\Se{S_{\text{e}}}

\newcommand{\journalfont}{\rm}  % this allows redefinition of the font later
\newcommand{\jou}[1]{{\journalfont #1\ }}
\newcommand{\joudef}[2]{\newcommand #1{\jou{\ignorespaces #2}}}

\newcommand{\eprintstyle}[1]{\textsf{#1}} % style for e-print references

\newcommand{\onlineversion}[1]{(related online version: \eprintstyle{#1})}
\joudef{\APAHA}  { Acta Physica Acad.~Sci.~Hung.}
\joudef{\ajp}    { Am.~J.~Phys.}
\joudef{\aaa}    { Astron.\ Astrophys.}
\joudef{\aip}    { Adv.\ Phys.}
\joudef{\adm}    { Adv.\ Math.}
\joudef{\am}     { Ann.\ Math.}
\joudef{\apb}    { Ann.\ Phys.\ (Berlin)}
\joudef{\apny}   { Ann.\ Phys.\ (N.Y.)}
\joudef{\apjs}   { Astrophys.\ J.\ Suppl.}
\joudef{\BAPMA}  {Bull.~Acad.~Pol.~Sci.~Ser.~Sci.~Math.~Astron.~Phys.}
\joudef{\baps}   { Bull.~Am.~Phys.~Soc.}
\joudef{\cjp}    { Can.\ J.\ Phys.}
\joudef{\cmda}   { Celest.\ Mech.\ Dyn.\ Astron.}
\joudef{\cmp}    { Commun.\ Math.\ Phys.}
\joudef{\cqg}    { Class.\ Quantum Grav.}
\joudef{\faa}    { Funct.\ Anal.\ Appl.}
\joudef{\grg}    { Gen.\ Rel.\ Grav.}
\joudef{\ijmpd}  { Int.\ J.\ Mod.\ Phys.\ D}
\joudef{\ijtp}   { Int.\ J.\ Theor.\ Phys.}
\joudef{\invm}   { Invent.\ Math.}
\joudef{\jm}     { J.\ Math.}
\joudef{\jmp}    { J.\ Math.\ Phys.}
\joudef{\jpa}    { J.\ Phys.\ A}
\joudef{\jpg}    { J.\ Phys.\ G\relax}
\joudef{\jpamg}  { J.\ Phys.\ A:\ Math.\ Gen.}
\joudef{\jpdap}  { J.\ Phys.\ D:\ Appl.\ Phys.}
\joudef{\lrr}    { Living Rev. Relativity}
\joudef{\mnras}  { Mon.\ Not.\ R.\ Ast.\ Soc.}
\joudef{\mpla}   { Mod.\ Phys.\ Lett.\ A} 
\joudef{\nature} { Nature}
\joudef{\nc}     { Nuovo Cim.}
\joudef{\npb}    { Nuc.\ Phys.\ B}
\joudef{\ph}     { Physica}
\joudef{\PHSTB}  { Phys.~Scripta}
\joudef{\pla}    { Phys.\ Lett.\ A}
\joudef{\plb}    { Phys.\ Lett.\ B}
\joudef{\pr}     { Phys.\ Rev.}
\joudef{\prep}   { Phys.\ Rep.}
\joudef{\PANUE}  {Phys.~Atom.~Nucl.}
\joudef{\pnas}   { Proc.\ Natl.\ Acad.\ Sci.\ USA}
\joudef{\prsla}  { Proc.\ Roy.\ Soc.\ Lond.\ A}
\joudef{\ptp}    { Prog.\ Theor.\ Phys.}
\joudef{\ptps}   { Prog.\ Theor.\ Phys.\ Suppl.}
\joudef\spj      { Sov.\ Phys.\ JETP}
\joudef\jetpl    { JETP Lett.}

\newcommand{\einsteinmaxwell}{04.40.Nr}  % Einstein-Maxwell spacetimes
\newcommand{\electrons}{14.60.Cd} % Electrons
\begin{document}
\bibliographystyle{prsty}

\setcounter{page}{1}
\title[]{SOME CONSEQUENCES OF GRAVITATIONALLY INDUCED \\ ELECTROMAGNETIC EFFECTS IN MICROPHYSICS
% \\[10pt] Working paper for publication in \\ Journal of the Korean Physical Society
}
\author{Kjell \surname{Rosquist} \\}
%\email{kr@physto.se}
%\thanks{}
\affiliation{Department of Physics, Stockholm University,
             106 91 Stockholm, Sweden \\ and \\
             ICRANet, Piazzale della Repubblica, 10 \\
             65100 Pescara, Italy}
%\date[]{Received January 5 2004}

\begin{abstract}

We discuss the relation between the gravitational and electromagnetic fields as governed by the Einstein-Maxwell field equations. It is emphasized that the tendency of the gravitational field to induce electromagnetic effects increases as the size of the system decreases. This is because the charge-to-mass ratio $Q/M$ is typically larger in smaller systems. For most astrophysical systems, $Q/M$ is $\ll 1$ while for a Millikan oil drop, $Q/M \sim 10^6$. Going all the way down to elementary particles, the value for the electron is $Q/M \sim 10^{21}$. For subatomic systems there is an additional phenomenon which comes into play. In fact, according to general relativity, the gravitational field tends to become dominated by the spin at distances of the order of the Compton wavelength.
The relevant quantity which governs this behavior is the ratio $S/M^2$ where $S$ is the (spin) angular momentum. For an electron, $S/M^2 \sim 10^{44}$. 
As a consequence, the gravitational field becomes dominated by gravitomagnetic effects in the subatomic domain. 
This fact has important consequences for the electromagnetic fields of spinning charged particles. 
To analyze this situation we use the asymptotic structure in the form of the multipole fields. 
Such an approach avoids the pitfalls should one try to use a near-field approach using some kind of semi-classical formulation of the Einstein-Maxwell equations for example. To obtain more exact results however, one must take quantum effects into account including radiative contributions. Although such effects are not considered in this work, the order of magnitude of the considered effects are not expected to change drastically when going to a quantum mechanical treatment. The most relevant solution of the Einstein-Maxwell equations in this context is the Kerr-Newman metric. It is the preferred solution which is in accord with all the four known multipole moments of the electron to an accuracy of one part in a thousand. Our main result is that general relativity predicts corrections to the Coulomb field for charged spinning sources. Experimentally verifiable consequences include a predicted electric quadrupole moment for the electron, possible quasi-bound states in positron-heavy ion scattering with sizes corresponding to observed anomalous peaks, as well as small corrections to energy levels in microscopic bound systems such as the hydrogen atom.

\end{abstract}

\pacs{\einsteinmaxwell, \electrons}

\keywords{General relativity, electromagnetism, electrons}

\maketitle

%%%%%%%%%%%%%%%%%%%%%%%%%%%%%%%%%%%%%%%%%%%%%%%%%%%%%%%%%%%%%%%%%%%%%%%%%%%%%%
\section{Introduction}
The main purpose of this paper is to discuss how the values of the electromagnetic and gravitational multipole parameters of an isolated system affect the asymptotic structure of the electromagnetic and gravitational fields via the Einstein-Maxwell field equation. Although the systems we have in mind may be microscopic, we adopt a purely classical point of view, since we are only dealing with the asymptotic structure in the form of the multipole fields. In this way we avoid the problems which would arise in a near-field approach using some kind of semi-classical formulation of the Einstein-Maxwell equations for example. To obtain more exact results, one must obviously take quantum effects into account including radiative contributions. Our analysis indicates that the Einstein-Maxwell field equations can severely constrain the form of the multipole structure because of non-linear effects. In particular, this has implications for elementary particles such as the electron.

It is usually stated that gravitational effects are irrelevant at the nuclear scale. This conclusion is drawn from a comparison between the Coulomb and the (Newtonian) gravitational forces between two protons. It is certainly true that the ratio between the gravitational ($f_\text{N}$) and Coulomb ($f_\text{C}$) forces is negligbly small, $f_\text{N}/f_\text{C} \sim 10^{-36}$. However, as discussed below, this is not the whole story. There is another relevant comparison which should be made. This is the relation between the field strengths. The forces depend not only on the fields, but also on the charges involved. However, the fields may well balance each other even though the forces do not. As it turns out, a meaningful comparison between forces and fields which remains valid down to subatomic scales requires a general relativistic analysis. The reason is that in general relativity, both the static (gravitoelectric) and the stationary (gravitomagnetic) moments function as sources of the gravitational field. In particular, the gravitomagnetic field corresponding to the spin of an electron, for example, becomes comparable to the gravitoelectric field at the Compton scale. This in turn induces electromagnetic effects via the Einstein-Maxwell equations. These effects lead to corrections to both the Coulomb and the magnetic fields of the electron. The most prominent signature is that the electron will acquire an electric quadrupole \cite{Rosquist:2006}. The corrections can in principle be measured in low energy ($\text{keV}$) scattering experiments. In fact, the size of the effect coincides with anomalous peaks observed in scattering of positrons against heavy ions \cite{Greiner&Reinhardt:1995, Sakai_etal:1993, Bargholtz_etal:1987, Erb_etal:1986}. Corrections to the Coulomb force may also influence the energy levels of the hydrogen atom. Such effects have been discussed by Pekeris and Frankowski \cite{Pekeris&Frankowski:1989} and by Gair \cite{Gair:2002}.

%%%%%%%%%%%%%%%%%%%%%%%%%%%%%%%%%%%%%%%%%%%%%%%%%%%%%%%%%%%%%%%%%%%%%%%%%%%%%%
\section{Newtonian gravity and electromagnetism}
%
%----------------------------------------------------------------------------%
\subsection{The classical unification scale}
\begin{figure}[t!]
\includegraphics[width=13.0cm]{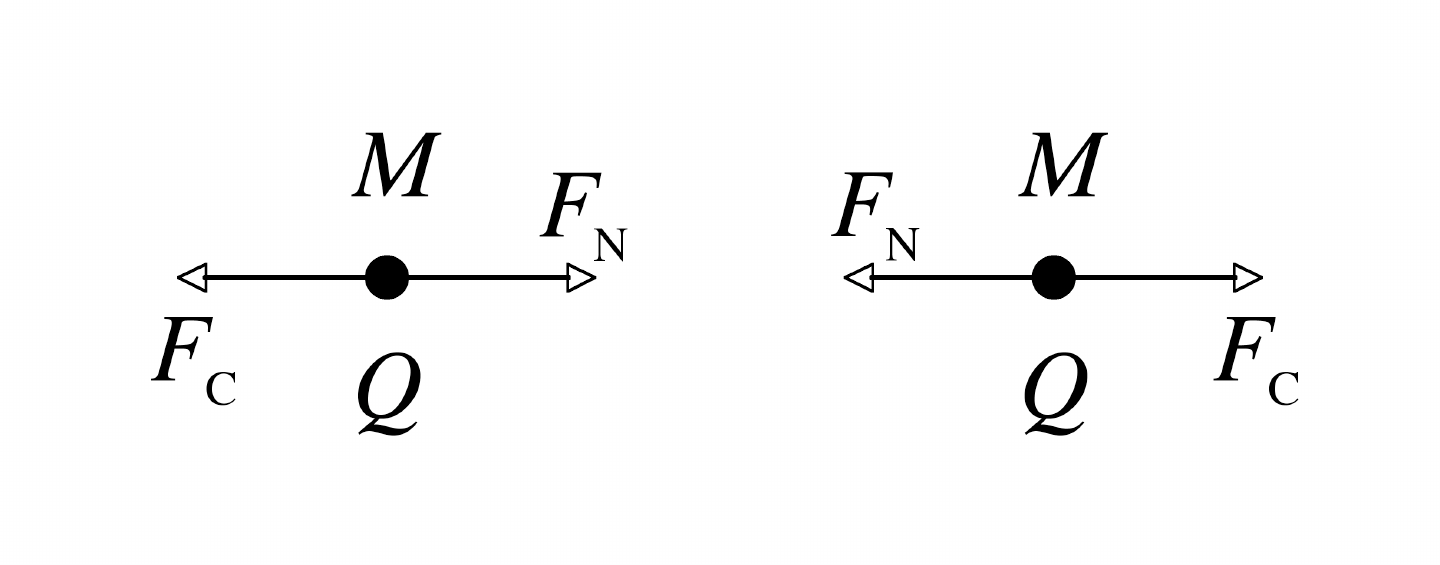}
\caption{Balanced gravitational and electric forces. If $Q=e$ (the elementary charge), then what is $M$?} \label{balanced_forces}
\end{figure}

We use geometric units \cite{Misner_etal:1973} in which the speed of light and Newton's gravitational constant are set to unity ($c=1$, $G=1$) and the electric permittivity is set to $\epsilon_0 = (4\pi)^{-1}$. Then Newton's law of gravity and the Coulomb force take the forms
\begin{equation}
   f_{\text{N}} = \frac{M_1M_2}{r^2} \ ,\qquad
   f_{\text{C}} = \frac{Q_1Q_2}{r^2} \ .
\end{equation}
As a consequence, charge and mass have the same dimension in these units. This means that it makes sense to compare the numerical values of masses and charges if one uses these standard units. Let us consider a gedanken experiment in which two charged particles are balanced by gravitational and electromagnetic forces as illustrated in Fig.\ref{balanced_forces}. The particles are assumed to be identical with masses $M_1=M_2=M$ and charges $Q_1=Q_2=Q$. To achieve balance we require that Newton's gravitational force $f_{\text{N}}$ has the same magnitude as Coulomb's force $f_{\text{C}}$, that is $|f_{\text{N}}| = |f_{\text{C}}|$. To be more specific, let us assume that $Q=e$ where $e$ is the elementary charge. We then adjust the mass $M$ to the value for which the forces are balanced. This gives the Stoney mass \footnote{The Irish physicist G.J.~Stoney first suggested $\mS=e$ as a unit of mass in 1881. Note that our emphasis in this context is the near equality between the Stoney mass and the Planck mass. As neither is nowhere near the observed masses of elementary particles, one should be aware that to consider either as a fundamental mass scale amounts to an enormous extrapolation. In fact, even the notion of a fundamental mass scale may be irrelevant. After all, the basic theories of physics, electromagnetism and general relativity are invariant under a rescaling of the mass.} \cite{Stoney:1881,Okun:2002,Okun:2004} $M = \mS = e \approx 2\, \mu\text{g}$ (where we have temporarily reverted to conventional units). It is only one order of magnitude lower than the Planck mass $m_{\text{P}} = \sqrt{\hbar} \approx 20\,\mu \text{g}$. The ratio between them is given by the square root of the fine structure constant, $\mS / m_{\text{P}} =  \alpha^{1/2} = \sqrt{e^2/\hbar} \sim 10^{-1}$. We remark in passing that the approximate equality between the Planck mass and the elementary charge, $m_{\text{P}} \approx 0.1 \, e$, could be more than a coincidence. There is another way of viewing the above thought experiment. The mass scale $\mS$ can clearly be considered as the scale of unification of gravity and electromagnetism at the classical (macroscopic) level. The fact that this scale is so close to the conjectured unification \cite{Wilczek:2005} scale $m_{\text{P}}$ of the fundamental forces at the quantum (microscopic) level provides a different perspective on the connection between macroscopic (gravity and electromagnetism) and microscopic (weak and strong interactions) physics. Whatever the interpretation, the above argument shows that classical Newtonian gravity combined with the Coulomb force leads in a natural way to a mass scale which coincides with the Planck mass scale to within an order of magnitude.

%----------------------------------------------------------------------------%
\subsection{The relative strength of gravitational and electromagnetic forces}
When discussing the relative strength of gravity and electromagnetism, the focus in the past has always been on the relation between the forces. However, it is of paramount importance in this context to consider also the relation between the fields. This is one of the main points of this contribution. In fact, the relation between the fields should be considered as more fundamental since it is governed only by the field equations themselves. The relation between the forces, on the other hand, depends not only on the fields but also on the charges. In order to obtain a clear separation between these two aspects, we start by considering the ratios between the forces and then proceed in the next section to discuss the relation between the fields.
Taking the quotient of the electromagnetic and gravitational forces gives
\begin{equation}
   \frac{f_{\text{C}}}{f_{\text{N}}}
    = \frac{Q_1}{M_1} \cdot \frac{Q_2}{M_2} \ .
\end{equation}
It follows from this relation that the relative strength of gravitational and electric forces depends on the quotient $Q/M$
\footnote{In a general relativistic analysis, this statement is not strictly true for the short distance interactions to be discussed in the following sections.}.
That is to say that when we measure $Q$ and $M$ in comparable units as discussed above, then the relative strength depends on the numerical values of the charges and the masses rather than attributing it solely to a difference in coupling constants as is usually done. It is therefore of interest to study the dimensionless ratio $Q/M$ for various physical systems. Starting from the largest scales, astrophysical objects typically have $Q/M \ll 1$. Consider next desktop size objects ($\sim 1\, \text{cm}$). We can easily observe electrostatic forces by the bare eye, but gravitational forces between desktop objects are too weak to be observed in that way. This shows that we can have $Q/M>1$ in this regime. Going down  further in size to about $10^{-4} \, \text{cm}$, a Millikan oil drop has $Q/M \sim 10^6$. Continuing all the way down to the microscopic regime, an electron has $Q/M \sim 10^{21}$. The above discussion thus indicates that the charge-to-mass ratio tends to become larger for smaller systems.

The charge and mass represent the lowest multipole moments of the electric field and its gravitational counterpart, namely the Newtonian gravitational field, or the gravitoelectric field in general relativity. The lowest magnetic multipole, on the other hand, is the magnetic dipole moment due to the absence of magnetic monopoles. The gravitational counterpart of the magnetic dipole is the angular momentum (orbital or spin) or gravitomagnetic dipole moment.
In Newtonian gravity, the gravitomagnetic multipoles appear only as inertial contributions to the kinetic energy. In general relativity on the other hand, the gravitomagnetic multipoles also act as sources of the gravitational field in analogy with the magnetic part of the electromagnetic field. The lowest gravitomagnetic multipole is the angular momentum (orbital or spin) which is the gravitational analogue of the magnetic dipole (see also \cite{Rosquist:2007a}).

%%%%%%%%%%%%%%%%%%%%%%%%%%%%%%%%%%%%%%%%%%%%%%%%%%%%%%%%%%%%%%%%%%%%%%%%%%%%%%
\section{BALANCE BETWEEN CURVATURE AND ELECTROMAGNETIC FIELD}
Having considered the relation between the forces in the previous section, we now come to the relation between the fields. In particular, we consider the relation between the strengths of the curvature and the electromagnetic field. We are assuming that gravity and electromagnetism obey the Einstein-Maxwell equations in the classical regime. The field equations are
\begin{equation}
   R^\mu{}_\nu = 8\pi T^\mu{}_\nu
               = 8\pi (-F^\mu{}_\lambda F^\lambda{}_\nu
    + \textstyle\frac14 \delta^\mu_\nu F^\lambda{}_\sigma F^\sigma{}_\lambda)
\end{equation}
where  $R^\mu{}_\nu$ is the Ricci tensor, $T^\mu{}_\nu$ is the stress-energy tensor and $F^\mu{}_\nu$ is the Maxwell field. This shows that generically there is a balance between the curvature and electromagnetic field which we can write symbolically in the form
\begin{equation}\label{Einstein-Maxwell_symbolic}
   R \sim F^2 \ .
\end{equation}
The Ricci tensor is given by the expression
\begin{equation}
   R_{\mu\nu} = 2\Gamma^\alpha{}_{\mu[\nu,\alpha]}
         +2\Gamma^\beta{}_{\mu[\nu} \Gamma^{\alpha}{}_{\alpha]\beta}
\end{equation}
which can be written in the symbolic form
\begin{equation}\label{Ricci_symbolic}
   R \sim \partial_x \Gamma + \Gamma^2 \ .
\end{equation}
The gravitational and electromagnetic forces on a test particle with mass $m$ and charge $q$ are given symbolically by
\begin{equation}\label{forces}
   f_{\text{grav}} \sim m \Gamma \ ,\qquad f_{\text{em}} \sim q F 
\end{equation}
where the first relation can be considered as a symbolic form of the geodesic equation and the second relation as a symbolic form of the Lorentz force equation. Therefore we can regard the connection $\Gamma$ as the gravitational field in the sense of being the force per unit mass, $\Gamma \sim f_{\text{grav}} /m$, as analogously the electromagnetic field is the force per unit charge, $F \sim f_{\text{em}} /q$. Since the fields $\Gamma$ and $F$ have the same dimension they can be directly compared. 
From (\ref{Einstein-Maxwell_symbolic}) and (\ref{Ricci_symbolic}) we see that generically the relation between the gravitational and electromagnetic fields has the symbolic form 
\begin{equation}
   \Gamma \sim F \ .
\end{equation}
The meaning of this relation is that generically the gravitational and electromagnetic fields balance each other. In the next section we examine how this balance works for an electrically charged spherically symmetric system.

\subsection{The Reissner-Nordstr\"om solution}
In general relativity there is a unique solution in the case of a spherically symmetric electrovacuum field. It is known as the Reissner-Nordstr\"om metric \cite{Misner_etal:1973}.
The Riemann tensor for the Reissner-Nordstr\"om field has two independent components, one in the Ricci part and one in the Weyl part. Modulo numerical factors, they are given by (in a certain orthonormal frame)
\begin{equation}
   R_{\mu\nu} \sim \frac{Q^2}{r^4} \ ,\qquad
   C_{\mu\nu\kappa\lambda} \sim -\frac{2M}{r^3} + \frac{Q^2}{r^4} \ . 
\end{equation}
The electromagnetic field has the Coulomb form
\begin{equation}
   F_{\mu\nu} \sim \frac{Q}{r^2}
\end{equation}
leading to a stress-energy tensor having the magnitude
\begin{equation}
   T_{\mu\nu} \sim (F_{\mu\nu})^2 \sim \frac{Q^2}{r^4} \ .
\end{equation}
These relations show how the Ricci part of the gravitational field is everywhere in balance with the electromagnetic field. In some sense this is a trivial consequence of the equality between the left and right hand sides of the Einstein-Maxwell equations. We are displaying this balance explicitly here to emphasize its relevance for microscopic physics as will be clearly seen in the next section when the spin is also taken into account. In this spherically symmetric setting, the electromagnetic field retains its macroscopic Coulomb form also in the microscopic domain. However, the spin will bring in drastic changes to this picture as we will see shortly.

Comparing now the Ricci and Weyl curvatures, we see that the Weyl part dominates the gravitational field in the asymptotic region $r \gg Q^2/M$ while the Ricci part dominates in the near zone $r \ll Q^2/M$. Asymptotically, the Weyl tensor represents the Newtonian limit corresponding to the second derivative of the gravitational potential $V'' \sim M/r^3$. Therefore, for the electron, the Newtonian region corresponds to $r \gtrsim r_{\text{class}}$ where $r_{\text{class}} = e^2/m_{\text{e}}$ is the classical electron radius. It is related to the reduced Compton wavelength $\lambdabar_{\text{C}}$ by, $r_{\text{class}} = \alpha \lambdabar_{\text{C}} \ll \lambdabar_{\text{C}}$ where $\alpha$ is the fine structure constant. Hence, we see that the Newtonian region extends well inside the quantum regime. Although we cannot draw any firm conclusions from this result, it is nevertheless a warning that gravity may come into play at scales which are much larger than the Planck length, $\ell_{\text{P}} \sim 10^{-20} r_{\text{class}}$. As we will soon see, more dramatic effects appear when the spin is taken into account.

The balancing of the gravitational and electromagnetic field was noted long ago by Johnston, Ruffini and Zerilli \cite{Johnston_etal:1973, Johnston_etal:1974}. In their first paper they showed that a neutral particle falling into a Reissner-Nordstr\"om black hole generates electromagnetic radiation. They also calculated the flux of electromagnetic radiation and found that it was of the same order of magnitude as the flux of gravitational radiation. This is an example of a gravitationally induced electromagnetic effect. In the second paper they considered the same situation but with a charged particle which provides an example of the opposite effect, namely electromagnetically induced gravitational radiation. Earlier, Gertsenshtein discussed briefly the possibility of resonance between electromagnetic and gravitational radiation \cite{Gertsenshtein:1962}.
%Such a resonance is now proposed as an efficient mechanism for detecting 
%gravitational radiation.

%%%%%%%%%%%%%%%%%%%%%%%%%%%%%%%%%%%%%%%%%%%%%%%%%%%%%%%%%%%%%%%%%%%%%%%%%%%%%%
\subsection{Asymptotic properties of the electron}
As discussed in the previous section, the Reissner-Nordstr\"om solution cannot be applied at the classical level to the electron's gravitational field in the region of Ricci domination due to quantum effects setting in. In this section we consider also the spin and the magnetic moment of the electron. It is well-known that mass, spin, charge and magnetic dipole comprise all the four known multipole moments of the electron. Since both the spin and the magnetic moment break spherical symmetry, the Reissner-Nordstr\"om Einstein-Maxwell field cannot adequately describe the asymptotic ($r \rightarrow \infty$) fields of the electron, i.e.\ the multipole structure. To take into account also the spin and the magnetic dipole one must use a solution of the Einstein-Maxwell equations which carries those additional moments. By far the simplest is the Kerr-Newman solution \cite{Misner_etal:1973}. Apart from being the simplest solution, the Kerr-Newman field has a number of other properties which makes it the prime candidate for the asymptotic field of the electron. Starting with the magnetic moment and the spin, the Kerr-Newman solution has precisely the right $g$-factor ($g=2$) to allow for the ratio between the spin and the magnetic moment (except only for the small $\sim 10^{-3}$ radiative corrections to the $g$-factor). Moreover, the very fact that the Kerr-Newman solution has $g=2$ shows that its angular momentum is a spin rather than an orbital angular momentum. The Kerr-Newman solution reduces to the Reissner-Nordstr\"om solution in the limit of small spin. Also, the Kerr-Newman solution is the unique axisymmetric Einstein-Maxwell field which admits a pair of conserved quantities which can serve as generalizations of the pair ($J^2, J_z$), something which is needed for a standard quantum mechanical treatment (cf.\ the works by Pekeris and Frankowski \cite{Pekeris:1989} and by Gair \cite{Gair:2002}). Another striking characteristic of the Kerr-Newman solution is the fact that the corresponding electromagnetic Lagrangian is finite \cite{Rosquist:2007b}. As is well-known the Lagrangian for the Coulomb field is divergent. Indeed, any finite superposition of electromagnetic moments leads to a divergent Lagrangian. However, when the spin parameter is nonzero, the electromagnetic Lagrangian has the value zero, $L_{\text{EM}}=0$. Taking the limit, $a \rightarrow 0$ therefore does not give the correct value for $L_{\text{EM}}$. This is one instance of the non-perturbative character of the Kerr-Newman field.

%See \cite{Rosquist:2007--Kerr-Newman representations} for another example of %a non-perturbative aspect of the Kerr-Newman solution.

Although there seems to be no controversy in using the Reissner-Nordstr\"om solution as a first approximation for the far field of the electron, the adoption of the spin $S_\text{e} =\hbar/2$ as a classical parameter by setting the Kerr-Newman parameter to $a_\text{e} = S_\text{e}/ m_\text{e} = \hbar/2 m_\text{e}$ may call for some additional motivation. There is a rather widely held view (even in textbooks) that the spin of an elementary particle is a pure quantum phenomenon. However, spin angular momentum is not only quantum mechanical. Actually, even within quantum mechanics, spin cannot be separated from the orbital angular momentum. As is well-known, only the total angular momentum operator $\vec{J}= \vec{L} + \vec{S}$ commutes with the Dirac Hamiltonian, while separately the spin and orbital angular momenta do not (see any textbook on relativistic quantum mechanics, e.g.\ \cite{Gross:1999}).
It is also a standard procedure to decompose the angular momentum of the classical electromagnetic field in two parts, one which depends on the position and one which is independent of the position. The two parts are interpreted as the orbital and spin angular momentum respectively (see e.g.\ \cite{Gross:1999}).
In particular, black holes do not carry orbital angular momentum, they have purely spin angular momenta. This is the lesson we learn from the $g=2$ value of the $g$-factor. For further discussion of gravity and spin in the microscopic domain, see \cite{Singh_etal:2004,Singh&Mobed:2007,Bini&Lusanna:2007}.

The Kerr-Newman solution represents a black hole if and only if the inequality $Q^2+a^2 \leq M^2$ holds. The limiting case $Q^2 + a^2 = M^2$ is known as an extremal black hole. Systems satisfying $M^2 < Q^2 + a^2$ are often referred to as overextreme (or hyperextreme). It deserves to be noted however, that there is nothing physically extreme about systems with $M<a$. On the contrary, many physical systems have $a>M$. To take one example, the solar system has $a/M \approx 40$. Another example is the spinning disk in a CD player which has $a/M \sim 10^{19}$. In fact, examining systems of different sizes, one finds that the dimensionless ratio $a/M$ is typically larger in smaller systems \cite{Rosquist:2007a}. It is therefore more to the point to consider the limiting case $Q^2+a^2 =M^2$ as a critical surface in parameter space. Since the case $M^2< Q^2+a^2$ contains desktop physics examples as well as microscopic physics, it is naturally viewed as the \emph{subcritical} case. Black holes should then consequently be referred to as \emph{supercritical} systems.

The four known gravitational and electromagnetic multipole moments of the electron expressed in geometric units are: the mass $\me$, the spin $\Se= \hbar/2$, the charge $e$ and the magnetic moment which is given by $\mu= (e/\me) S = e\hbar / (2\me) = e a_\text{e}$ where $a_\text{e}= S/\me = \hbar/(2\me)$. The spin is a gravitomagnetic dipole moment \cite{Mashhoon:2000, Mashhoon&Kaiser:2006}, i.e.\ a gravitational analogue of the magnetic dipole moment. The values of the electron's parameters imply the strong inequalities $a_\text{e} \gg e \gg \me$ \cite{Rosquist:2006}. The corresponding Kerr-Newman field is therefore of the subcritical (i.e.\ overextreme) type and is dominated by the spin in the near zone. In particular, it has no horizon and, as is readily shown, it has no ergoregion. An important conclusion is that gravity tends to become spin dominated in the subatomic domain.

The Kerr-Newman solution can be written in the form \cite{Rosquist:2007c}
\begin{equation}
   g_{\text{K}} = -h(r) (M^0)^2 + h(r)^{-1} (M^1)^2 + (M^2)^2 + (M^3)^2
\end{equation}
where
\begin{equation}
   h(r) = 1-\frac{2Mr-Q^2}{r^2+a^2}
\end{equation}
and $M^\mu$ is a certain orthonormal Minkowski frame (meaning that $\eta= \eta_{\mu\nu} M^\mu M^\nu$ is the Minkowski metric). The electromagnetic field in the orthonormal frame
\begin{equation}
   \chi^0 = h(r)^{1/2} M^0 \ ,\qquad \chi^1 = h(r)^{-1/2} M^1 \ ,\qquad
   \chi^2 = M^2 \ ,\qquad \chi^3 = M^3
\end{equation}
as expressed in Boyer-Lindquist coordinates is given by
\begin{equation}\label{KN_EM}
   F^\mu{}_\nu = \begin{pmatrix}
                   0 &  E &  0 &  0 & \\
                   E &  0 &  0 &  0 & \\
                   0 &  0 &  0 &  B & \\
                   0 &  0 & -B &  0 \end{pmatrix} \ ,\qquad
   \begin{cases}
    E = \displaystyle\frac{Q(r^2-a^2\cos^2\!\theta)}
                          {(r^2+a^2\cos^2\!\theta)^2} \\[9pt]
    B = \displaystyle\frac{2Qar\cos\theta}
                          {(r^2+a^2\cos^2\!\theta)^2} \ .
   \end{cases}
\end{equation}
It follows that the corresponding stress-energy tensor depends on the single function
\begin{equation}
   T^\mu{}_\nu \sim E^2 + B^2 = \frac{Q^2}{(r^2 + a^2\cos^2\!\theta)^2} \ .
\end{equation}
The Kerr-Newman Ricci tensor has a single independent component which is proportional to the square root of the invariant (see \cite{Chandrasekhar:1983} and \cite{Cherubini_etal:2002} for the curvature components and invariants)
\begin{equation}\label{Ricci_square}
   R^{\mu\nu} R_{\mu\nu} = \frac{4Q^4}{(r^2 + a^2\cos^2\!\theta)^4} \ .
\end{equation}
The Ricci tensor itself is again, as it should, proportional to the stress-energy tensor which exhibits explicitly the balance $R^\mu{}_\nu \sim (F^\mu{}_\nu)^2$ between the Ricci curvature and the electromagnetic field. In this case however, the electromagnetic field is not purely electric but has also a magnetic part. Its explicit form as given in \eqref{KN_EM} shows clearly that it has the Coulomb form at infinity while it deviates quite drastically from that form in the near zone. Indeed, apart from the nonzero magnetic part, the electric field itself differs from the Coulomb form. When analyzing the expressions in \eqref{KN_EM} one must be aware that the coordinates are not spherical but correspond to oblate spheroidal coordinates. Even bearing this in mind though, it is clear from the form of $E$ that it changes sign at the spheroidal radius $r= a\cos\theta$. Therefore the electric force on a test particle changes its nature from repulsive to attractive at that radius (or vice versa depending on the sign of the test particle) \cite{Rosquist:2006}.

It is now clear from \eqref{KN_EM} that general relativity predicts that the gravitational field of a spinning charge induces changes in its electromagnetic field which makes it deviate from a pure Coulomb form at small distances. As we have seen, the deviations from the Coulomb form occur at approximately the scale given by the spin parameter, $a$. For an electron, this corresponds to half the reduced Compton wavelength. At first sight, one might think that this would immediately invalidate general relativity at this level. However, the changes, in terms of measurable effects for an electron are quite subtle (see \cite{Rosquist:2006} for some more comments on this issue). One reason is that the effects appear only in the higher multipoles starting with the electric quadrupole. In fact, by regarding the Kerr-Newman solution as a prototypical solution of general relativity, it follows that spinning particles should have an infinite hierarchy of multipoles. In my view, one must take general relativity seriously also at this level and consider carefully the consequences. The prevailing view however, as is well known, is that general relativity does apply at the level of subatomic physics but that its effects are very small for sub-Planckian energies. However, the above arguments make it very plausible that general relativity does come in already at the Compton scale via its interaction with the electromagnetic field. In this context, it is also very important to understand that, even though the gravitational and electromagnetic fields are in balance, it doesn't mean that the \emph{forces} are equal. As emphasized above, the forces depend not only on the fields, but also on the ratios $Q/M$. Therefore, when the asymptotic gravitational and electromagnetic fields are modelled by the Kerr-Newman solution, the force between two electrons is still very much dominated by the electromagnetic interaction, even at the Compton scale (see \cite{Rosquist:2006} for numerical estimates).

The Kerr-Newman Weyl tensor has two independent components which can be expressed in terms of the Weyl spinor component $\Psi_2$ which has the real and imaginary parts
\begin{equation}\label{Weyl}
 \begin{split}
   \re \Psi_2 &= \frac{-r^2(Mr-Q^2) + (3Mr-Q^2)\,a^2\cos^2\!\theta}
                           {(r^2 + a^2\cos^2\!\theta)^3} \\[7pt]
   \im \Psi_2 &= \frac{[-r(3Mr-2Q^2)+Ma^2\cos^2\!\theta]\,a\cos\theta}
                           {(r^2 + a^2\cos^2\!\theta)^3} \ .
 \end{split}
\end{equation}
Here $\re\Psi_2$ represents the gravitoelectric part of the Weyl tensor while $\im\Psi_2$ represents the gravitomagnetic part. 
From these relations and \eqref{Ricci_square} it follows that the asymptotic behavior is given by (up to numerical factors)
\begin{equation}\label{curv_asymp}
   R_{\mu\nu} \sim \frac{Q^2}{r^4} \ ,\qquad
   \re\Psi_2 \sim \frac{M}{r^3} \ ,\qquad
   \im\Psi_2 \sim \frac{Ma\cos\theta}{r^4} \ .
\end{equation}
As in the Reissner-Nordstr\"om case, the Weyl tensor dominates in the asymptotic Newtonian regime. Note also that the dominating part of the curvature ($\sim M/r^3$) depends only on the mass. The spin and charge parameters do not contribute to the curvature for large $r$.
To estimate the curvature at the characteristic spin radius, $r=a$, we define a dimensionless radial coordinate by $\hat r= r/a$. The Weyl curvature then takes the form
\begin{equation}
 \begin{split}
   \re\Psi_2 &= \frac1{\ell^2} \cdot
     \frac{-\hat r^2(\hat r-\beta) + (3\hat r-\beta) \cos^2\!\theta}
          {(\hat r^2 + \cos^2\!\theta)^3} \\[7pt]
   \im\Psi_2 &= \frac1{\ell^2} \cdot 
     \frac{[-\hat r(3\hat r-2\beta)+\cos^2\!\theta]\cos\theta}
          {(\hat r^2 + \cos^2\!\theta)^3}
 \end{split}
\end{equation}
where
\begin{equation}
   \beta = \frac{Q^2}{Ma}
\end{equation}
and where the parameter
\begin{equation}
   \ell = \sqrt{\frac{a^3}{M}}
\end{equation}
can be interpreted as an approximate curvature radius whenever $\ell^2 |R_{\mu\nu \lambda\sigma}| \sim 1$.
Using the same parametrization for the Ricci curvature gives
\begin{equation}
  (R^{\mu\nu} R_{\mu\nu})^{1/2}
    = \frac1{\ell^2} \cdot \frac{\beta}{(\hat r^2 + \cos^2\!\theta)^2} \ .
\end{equation}
For the electron, we have $Q=e$ and $Ma= m_\text{e} a_\text{e} = \hbar/2$ so that $\beta=2\alpha \approx 0.015$. Setting $r=a_\text{e}$ and using $\beta \ll 1$ gives
\begin{equation}\label{curvature_compton}
 \begin{split}
   \re\Psi_2 &\approx \frac1{\ell^2} \cdot
                     \frac{3\cos^2\!\theta-1}{(1+\cos^2\!\theta)^3} \ ,\quad
   \im\Psi_2 \approx \frac1{\ell^2} \cdot 
     \frac{(\cos\theta-3)\cos\theta}
          {(1 + \cos^2\!\theta)^3} \\[7pt]
     (R^{\mu\nu} R_{\mu\nu})^{1/2} 
     &= \frac1{\ell^2} \cdot \frac{2\alpha}{(1 + \cos^2\!\theta)^2} \ .
 \end{split}
\end{equation}
This shows that $\ell$ is a rough estimate of the minimum curvature radius near $r=a_\text{e} = \lambdabar/2$. We can also draw the conclusion that the Weyl curvature is generically larger than the Ricci curvature at this scale because of the factor $\alpha$ in the expression for the Ricci curvature. Furthermore, it follows from \eqref{curvature_compton} that the gravitomagnetic curvature ($\im\Psi_2$) is of the same order of magnitude as the gravitoelectric curvature ($\re\Psi_2$) at the Compton scale. Even so, the gravitoelectric part depends on the spin parameter via $\ell$ and is therefore indirectly determined by the gravitomagnetic part.

To summarize this section we see that the departure from Newtonian gravity in the Kerr-Newman case occurs at $r \sim a = \lambdabar/2$, i.e.\ roughly the reduced Compton wavelength. At this scale, the curvature is determined in an essential way by the spin parameter, $a$, and is in this sense dominated by gravitomagnetic effects. We emphasize again that the gravitational forces are small in this regime. The only but significant role of the gravitational field is to induce electromagnetic effects via the Einstein-Maxwell equations.

%%%%%%%%%%%%%%%%%%%%%%%%%%%%%%%%%%%%%%%%%%%%%%%%%%%%%%%%%%%%%%%%%%%%%%%%%%%%%%
\section{CONCLUSIONS}
It follows from the Einstein-Maxwell field equations that the gravitational field can induce electromagnetic effects. In particular, we have pointed out that the classical field equations lead to a balance between the gravitational and electromagnetic fields at the microscopic level. When applied to the multipole parameters of the electron, this balance sets in at approximately the Compton scale. At this scale, the dominating source of the gravitational field is shifted from the mass to the spin. This is a non-Newtonian gravitomagnetic effect.

The fact that the electromagnetic forces (not fields!) dominate in the subatomic domain can be explained in terms of the large value of the ratio $e/m$ for elementary particles. Thus, even though the electromagnetic forces dominate over gravity in the microscopic domain, there must be a balance between the fields due to the Einstein-Maxwell equations. Although the analysis presented in this contribution is purely classical, it should be expected that while a full quantum mechanical treatment will modify the details of the interactions, the order of magnitude of the effects will remain the same.

Several arguments indicate that the Kerr-Newman solution is the strongest candidate to model the asymptotic multipole structure of the electromagnetic and gravitational fields of the electron. The predicted higher electromagnetic multipoles start with the electric quadrupole which should in principle be possible to measure in high precision low energy scattering experiments (see also \cite{Rosquist:2006}). The multipole structure may also be responsible for observed anomalous peaks in heavy ion scattering experiments \cite{Greiner&Reinhardt:1995, Sakai_etal:1993, Bargholtz_etal:1987, Erb_etal:1986}.

%%%%%%%%%%%%%%%%%%%%%%%%%%%%%%%%%%%%%%%%%%%%%%%%%%%%%%%%%%%%%%%%%%%%%%%%%%%%%%
\appendix

\begin{acknowledgments}

I wish to thank C.~Lozanovski and N.~Van den Bergh for helpful comments on a previous version of the manuscript.

\end{acknowledgments}

%============================================================================%
%\bibliography{kr}

%\begin{references}

%\end{references}

\newpage

\end{document}